\documentclass[12pt,letterpaper]{article}

\usepackage{natbib}
\usepackage{aas_macros}
\usepackage[dvips]{graphicx}
\usepackage{vmargin}

\setpapersize{USletter}
\setmarginsrb{1in}{1in}{1in}{1in}{5pt}{0mm}{5pt}{0mm}
\title{Fully Automated Approaches to Analyze Large-Scale Astronomy Survey Data}
\author{
	{\bf A. Pr\v sa} \\
	{\bf Villanova University \& University of Ljubljana} \\ 
	{\tt andrej.prsa@villanova.edu} \quad (610) 519-4887 \\
	\\
	with\\
	\\
	E.~F.~Guinan${}^1$, E.~J.~Devinney${}^1$, S.~G.~Engle${}^1$, M.~DeGeorge${}^1$, \\
	G.~P.~McCook${}^1$, P.~A.~Maurone${}^1$, J.~Pepper${}^2$, D.~James${}^3$, \\
	D.~H.~Bradstreet${}^4$, C.~R.~Alcock${}^5$, J.~Devor${}^5$, R.~Seaman${}^6$ \\
	T.~Zwitter${}^7$, K.~Long${}^8$, R.~E.~Wilson${}^{9}$, I.~Ribas${}^{10}$, A.~Gimenez${}^{11}$ 
}

\begin{document}

\maketitle

\begin{abstract}

Observational astronomy has changed drastically in the last decade: manually driven target-by-target instruments have been replaced by fully automated robotic telescopes. Data acquisition methods have advanced to the point that terabytes of data are flowing in and being stored on a daily basis. At the same time, the vast majority of analysis tools in stellar astrophysics still rely on manual expert interaction. To bridge this gap, we foresee that the next decade will witness a fundamental shift in the approaches to data analysis: case-by-case methods will be replaced by fully automated pipelines that will process the data from their reduction stage, through analysis, to storage. While major effort has been invested in data reduction automation, automated data analysis has mostly been neglected despite the urgent need. Scientific data mining will face serious challenges to identify, understand and eliminate the sources of systematic errors that will arise from this automation. As a special case, we present an artificial intelligence (AI) driven pipeline that is prototyped in the domain of stellar astrophysics (eclipsing binaries in particular), current results and the challenges still ahead.

\end{abstract}

\noindent
${}^1$Villanova University; ${}^2$Vanderbilt University; ${}^3$University of Hawaii; ${}^4$Eastern University; ${}^5$Harvard-Smithsonian CfA; ${}^6$NOAO; ${}^7$University of Ljubljana; ${}^8$STSCI; ${}^9$University of Florida; ${}^{10}$University of Barcelona; ${}^{11}$INTA/CSIC.

\newpage

\section{Introduction}

One of the most important changes in observational astronomy of the 21$^{\rm st}$ Century is a rapid shift from classical object-by-object observations to extensive automated surveys. As CCD detectors improve in sensitivity and their costs decrease, more and more small and medium-size observatories are refocusing their attention\footnote{A comprehensive list of more than a hundred such facilities may be found at {\tt http://www.astro.physik.uni-goettingen.de/\~{}hessman/MONET/links.html}.} to the investigation of stellar variability through systematic wide-field sky-scanning missions. This trend is additionally powered by the success of pioneering surveys such as EROS \citep{pd1998}, MACHO \citep{macho1995}, OGLE \citep{udalski1997}, ASAS \citep{pojmanski2002}, their space counterpart Hipparcos \citep{hipparcos1997} and others. Such surveys produce massive amounts of data that pose a significant challenge to reduction and analysis. Surveys and missions  currently commissioned (i.e. Kepler \citep{borucki2007}, LSST \citep{lsst2002}, Pan-STARRS \citep{kaiser2004} and Gaia \citep{perryman2001}) will produce petabytes of data daily; spectroscopic surveys such as RAVE \citep{steinmetz2006}, SEGUE \citep{newberg2003} and Hermes \citep{raskin2008} will open the doors for complementary spectroscopy for millions of sources. Yet currently-available tools fall short of the needs of proper analysis. % On the data front-end side, however, significant effort has been invested to propose, evaluate and implement rapid signal and data processing schemes that would be able to cope with the fire-hose of data such surveys will generate.

In this white paper we limit ourselves to stellar astrophysics (in particular, eclipsing binary stars -- EBs hereafter), but the points raised are readily applicable to other areas of astronomy, such as the study of pulsating variable stars, astroseismology, stellar rotation, population theory, etc.  To date, about a thousand papers have been published on EBs with physical and geometrical parameters determined to better than 3\% accuracy. For an eclipsing binary expert it takes 1-2 weeks to reduce and analyze a single eclipsing binary light-curve the old-fashioned way. There are currently about 10,000 photometric/RV data-sets that in principle allow modeling to a 3\% accuracy. According to Hipparcos results, about 0.8\% of the overall stellar population are EBs (917 out of 118,218 stars, \citealt{hipparcos1997}). Projecting these statistics to other large surveys allows estimating the number of EBs expected in survey databases: $\sim$136,000 in ASAS, $\sim$56,000 in the OGLE LMC field, $\sim$16,000 in OGLE SMC field, $\sim$80,000 in TASS \citep{droege2006}, etc. Gaia will make a revolution in these numbers since the aimed census of the overall stellar population is $\sim$1 billion down to $V = +20$\,mag \citep{perryman2001}, yielding millions of EBs and tens of millions of variable stars. Finally, with LSST essentially complete to $V=24.5$, the yield of EBs will reach the tens of millions. Even if all observational facilities collapsed at that point so that no further data got collected, it would still take 500 years for all 12,500 members of the IAU to analyze these data! Given the unique capability of EBs to yield accurate stellar masses, radii, temperatures and distances, and realizing that many of these are accessible by small-size ground instruments, EBs should definitely hold one of the top positions on observational candidates list.

\section{Brief review of process automation}

\begin{figure}[t]
\begin{center}
\includegraphics[width=8cm]{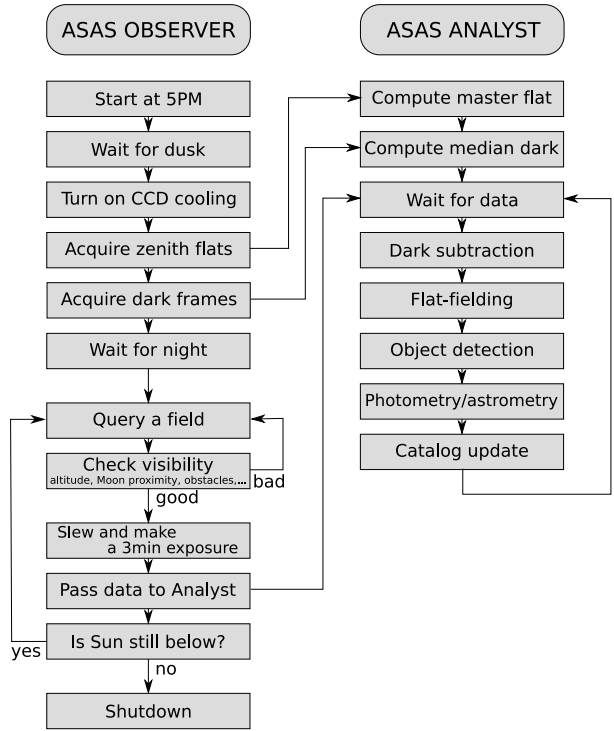} \\
\end{center}
\caption{\label{asas_pipeline} ASAS project's automated data acquisition pipeline \citep{pojmanski1997}.}
\end{figure}

Data acquisition is the most automated aspect of the pipeline. An example of a fully automatic data acquisition and analysis pipeline is that of the All-Sky Automated Survey (ASAS, \citealt{pojmanski1997}), depicted in Figure~\ref{asas_pipeline}. The level of sophistication is already such that it assures accurate and reliable data from both ground-based and space surveys.

\begin{figure}[t]
\begin{center}
\includegraphics[width=0.95\textwidth]{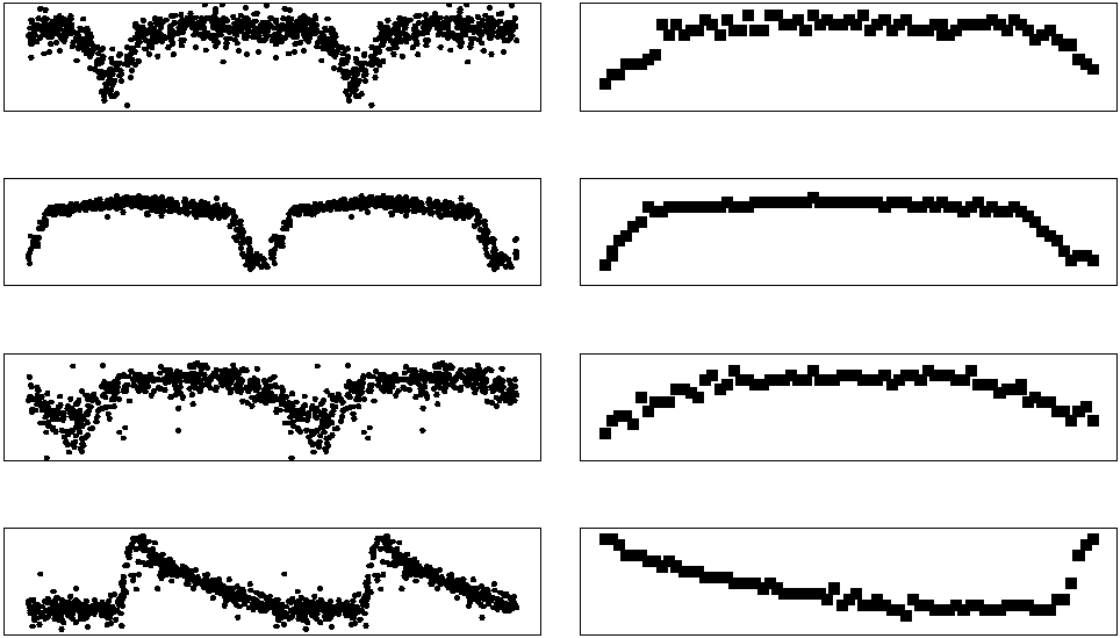}
\end{center}
\caption{\label{wyrzykowski} Conversion of phased light curves (left) to 70$\times$15 pix images (right), which are fed to the neural network image recognition algorithm. Taken from \citet{wyrzykowski2003b}.}
\end{figure}

Variability classification, however, has proved to be much more involved than perhaps initially expected. Fundamentally different objects (i.e.~radial pulsators and ellipsoidal variables) produce essentially indistinguishable light curves and follow-up spectroscopy is paramount for identifying their true nature. A series of systematic  analyses were conducted by \citet{rucinski1997a, rucinski1997b, rucinski1998} and later \citet{maceroni1999} and \citet{rucinski2001} that highlighted the importance of the Fourier decomposition technique (FDT) for classification of variable stars. The technique itself -- fitting a 4$^{\rm th}$ order Fourier series to phased data curves and mapping different types of variables in Fourier coefficient space -- was originally proposed for EBs  by \citet{rucinski1973} and has been used ever since, most notably for classifying ASAS data \citep{pojmanski2002, paczynski2006}. \cite{alcock1997}, analyzing 611 bright EBS from the MACHO database \citep{macho1995}, proposed a new decimal classification scheme for categorizing EB types. \citet{wyrzykowski2003b, wyrzykowski2004} identified 2580 EBs in the LMC and 1351 EBs in the SMC. They employed a novel classification approach  using Artificial neural networks (ANN) as an image recognition algorithm, based on phased data curves that have been converted to low-resolution images as depicted on Figure~\ref{wyrzykowski}. Their classification pipeline was backed up by visual examinations of results.

Approaches to automating light curve solutions have taken various forms to date. \citet{wyithe2001,wyithe2002}, in their work to establish the best distance indicators among detached and semi-detached binaries in the Small Magellanic Cloud, obtained starting parameters for the rigorous WD model by comparing each candidate light curve with a set of template model light curves, sending the best match to an automated version of the differential corrector program (DC). The latter could be computationally prohibitive to apply to the expected large future data-sets.  Employing less rigorous physical models, of course, is one approach to computational efficiency.  Thus, \citet{tamuz2006} employ the EBOP ellipsoidal model \citep{popper1981}.  Using this engine, they arrive at an initial solution after a combination of grid search, gradient descent and geometrical analysis of the LC. \citet{devor2005} illustrates an automated pipeline employing a simple model of spherical stars without tidal or reflection physics, whose starting values are similarly obtained. \citet{prsa2008} have devised a neural network based engine EBAI (Eclipsing Binaries via Artificial Intelligence; {\tt http://www.eclipsingbinaries.org}) that is capable of processing thousands of light curves in just a few seconds; it yields principal parameters of the analyzed variable.

\section{Artificial Intelligence}

\begin{figure}[b!]
\begin{center}
\includegraphics[width=0.8\textwidth]{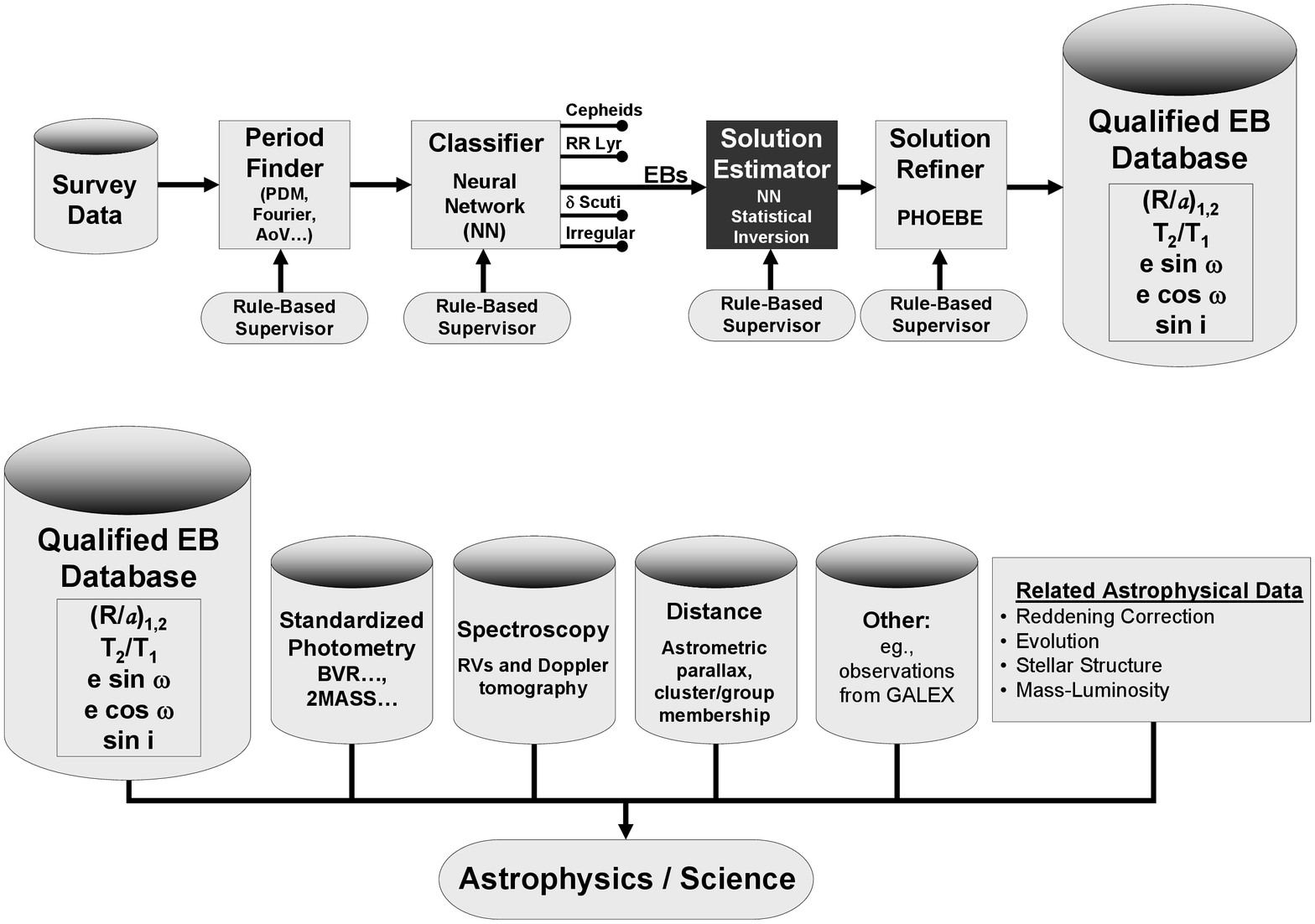} \\
\caption{Intelligent Data Pipeline (IDP). Complete survey data is piped through a period finder algorithm that is controlled by a rule-based system. All variable sources are then passed to the ANN-based classifier. Light curves consistent with EB signatures are passed to the Solution Estimator block.} \label{idp}
\end{center}
\end{figure}

Advances in Artificial Intelligence (AI) and the continued operation of Moore's law that predicts exponential growth of the processing power have created the opportunity for significant progress in solving the types of problems that are limited by the lack of human capital. A new approach, the \emph{Intelligent Data Pipeline} (IDP), has been prototyped in the domain of EBs which uses AI techniques to operate autonomously on large observational data-sets to produce results of astrophysical value. The IDP is designed to handle the complete process of variable discovery, classification of variability and management of the solution process for the discovered EBs (\citealt{devinney2005,devinney2006}; cf.~Fig.~\ref{idp}). The IDP employs ANNs in the processing modules, while the supervisory knowledge, now implicit in humans, is encoded in control modules as rules appropriate for each processing module. The supervisory modules have the task of keeping the process on track and providing physically meaningful results through each phase of the processing pipeline.

\begin{figure}[!b]
\begin{center}
\includegraphics[width=0.6\textwidth]{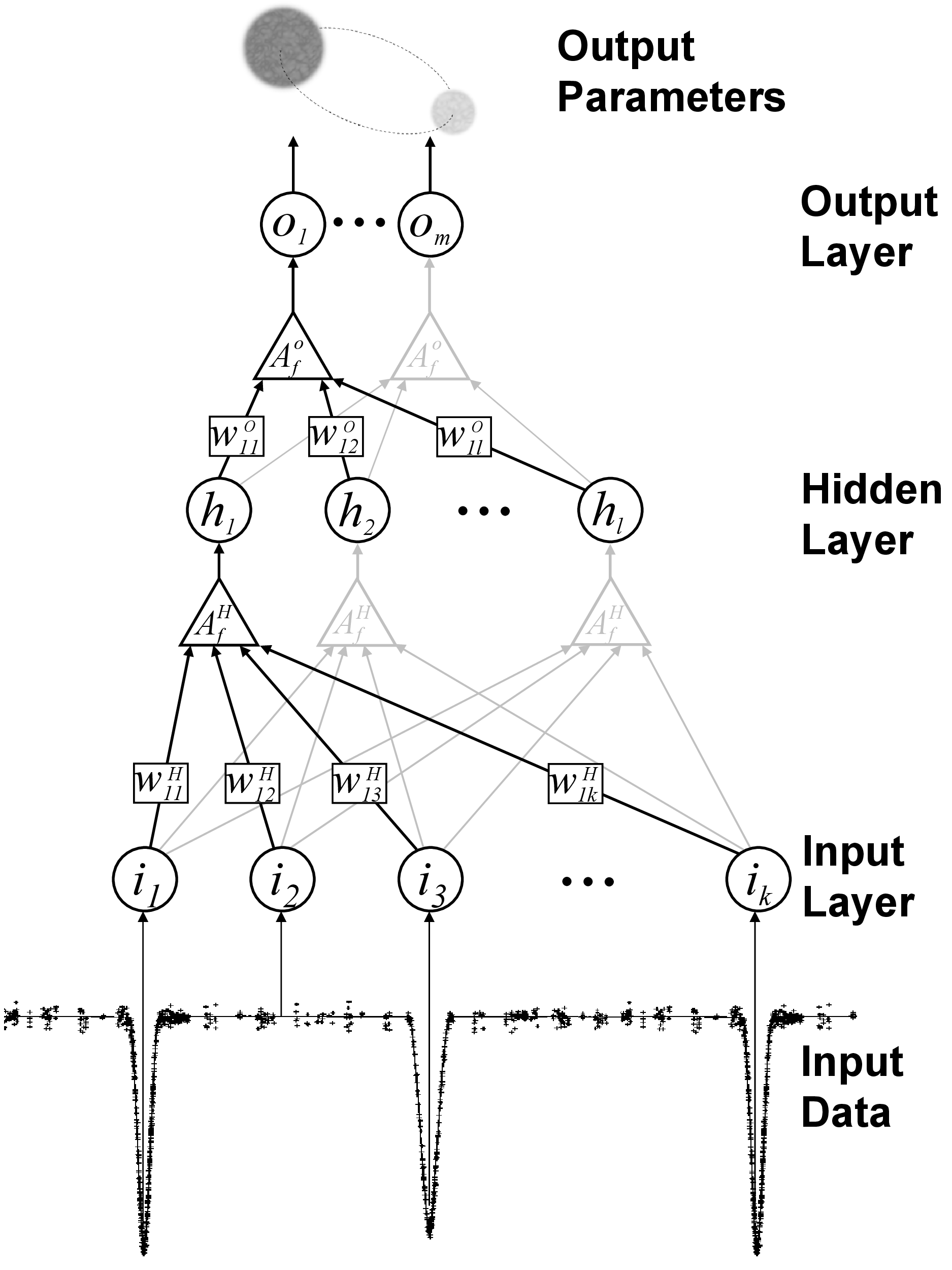} \\
\end{center}
\caption{Schematic view of a three-layered, fully connected ANN. \label{ann}}
\end{figure}

ANNs are very simple algorithms that  involve little beyond summation and multiplication, while having the capability of being \emph{trained} on a physical content.  While some may find the opaqueness of Artificial Neural Networks (ANN) problematic, their success in many areas, including classification, real-time robotic control and others is a powerful answer.

In their basic form, ANNs are systems of multiple layers (Fig.~\ref{ann}). Each layer consists of a given number of independent units. Each unit holds a single value. These values are propagated from each unit on the current layer to all units on the subsequent layer by weighted connections. Propagation is a simple linear combination $y_i = \sum_j w_{ij} x_j$, where $x_j$ are the values on the current layer, $w_{ij}$ are weighted connections, and $y_i$ are the values that enter the subsequent layer. Before they are stored in their respective units, $y_i$ are first passed through a (non-linear) \emph{activation function} $A_f$. This function, typically a sigmoid function -- $A_f (y_i) = 1/[1+\exp(-(y_j-\mu)/\tau)]$ -- introduces non-linear properties to the network. Coefficients $\mu$ and $\tau$ are selected so that $A_f (y_i)$ fall in the $(-1,1)$ interval. It is this value that is stored in the $i$-th unit on the subsequent layer. Layers in the three-layer network are usually denoted \emph{input}, \emph{hidden}, and \emph{output} layer. ANN is thus a non-linear \emph{mapping} from the input layer to the output layer. In our implementation in the domain of EBs, the ANN maps the input light curves to the output set of principal physical parameters.

Training the network implies determining the weights $w_{ij}$ on weighted connections. The back-propagation algorithm relies on a sample of LCs (the training set) with known physical parameters; these are called \emph{exemplars}. All LCs are propagated through the network and their outputs are compared to the known values. The weights are then modified so that the discrepancy between the two sets is minimized. This is an iterative process that needs to be done only once. After training, the network is ready to process any input LC extremely quickly.

\begin{figure}[!t]
\begin{center}
\includegraphics[height=\textwidth,angle=-90]{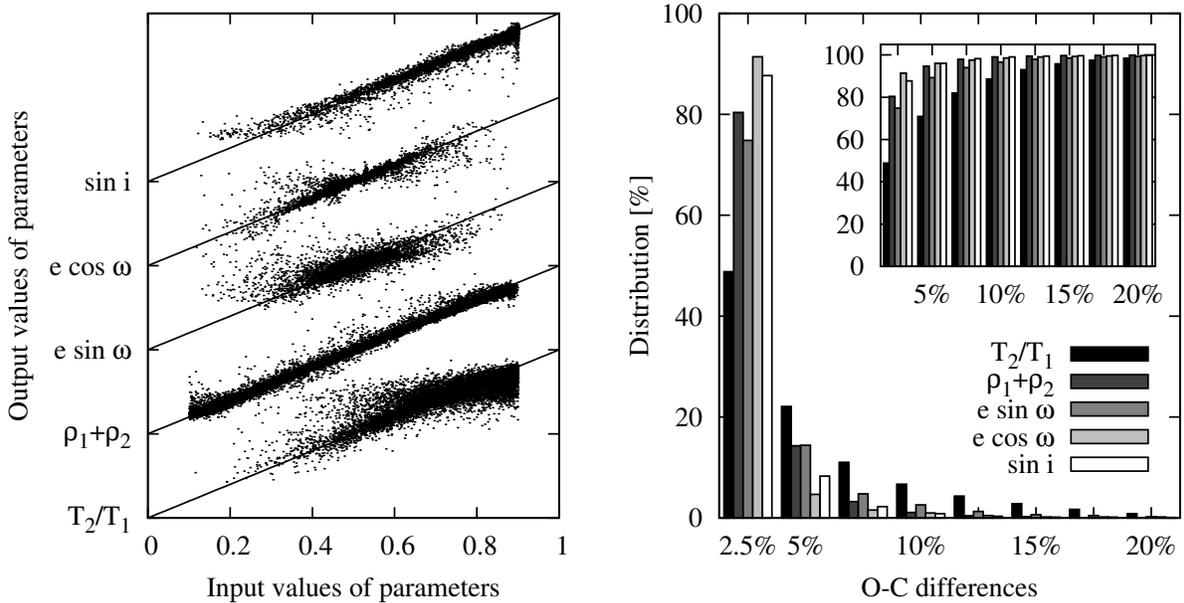} \\
\end{center}
\caption{ANN performance on a sample of 10,000 LCs. Left: comparison between the input parameters (known from generating a sample) and output parameters provided by the network. $T_{1,2}$ are effective temperatures of EB components, $\rho_{1,2}$ are their fractional radii, $e$ is eccentricity, $\omega$ is argument of periastron, and $i$ denotes inclination. Parameters are offset by 0.5 for clarity and a guideline is provided for easier comparison. Right: distribution of residuals (main graph) and their cumulative distribution (inset). The bars depict the fraction of EBs with errors between 0\% and 2.5\% (first bin), 2.5\% and 5\% (second bin), etc. 90\% of all LCs have errors smaller than 10\% in \emph{all} parameters.\label{res}}
\end{figure}

To evaluate ANN performance, we created a set of 10,000 synthetic light curves for eclipsing binary stars and passed it through a trained ANN. To each LC we added variable amounts of white noise, simulating different S/N ratios. Fig.~\ref{res} depicts the results that show clear statistical viability: 90\% of the sample resulted in parameters with errors less than 10\%. The success rate of recognition is comparable to that of the learning sample, and the underlying distribution of errors for both data-sets is indistinguishable. This demonstrates the capability of the ANN to successfully recognize data it has never seen before.

\section{Discussion}

The importance of results that will be achieved by developing novel fully automated approaches can hardly be overstated. In the domain of EBs, their analysis yields:
\begin{itemize}
    \item calibration-free physical properties of stars (i.e. masses, radii, surface temperatures, luminosities);
    \item accurate stellar distances;
    \item precise stellar ages;
    \item stringent tests of stellar evolution models.
\end{itemize}
 
The products of state-of-the-art EB modeling are seminal to many areas of astrophysics:
\begin{itemize}
    \item calibrating the cosmic distance scale;
    \item mapping of clusters and other stellar populations (e.g. star-forming regions, streams, tidal tails, etc) in the Milky Way;
    \item determining initial mass functions and studying stellar population theory;
    \item understanding stellar energy transfer mechanisms (including activity) as a function of temperature, metallicity and evolutionary stage;
    \item calibrating stellar color-temperature transformations, mass-radius-luminosity relationships, and other relations basic to a broad array of stellar astrophysics;
    \item studying stellar dynamics, tidal interactions, mass transfer, accretion, chromospheric activity, etc.
\end{itemize}

In addition, spectroscopic surveys such as RAVE, SEGUE and Hermes will provide observations of thousands of spectroscopic binaries that will allow the determination of metallicity, leading to chemical tagging, galactic stratigraphy and population memberships.

\bigskip

The enormous inflow of data that marked the previous decade exposed the deficiencies of the analysis tools in this decade. Manual analysis will have to be limited to the astrophysically most interesting cases; all other sources will need to be processed in a fully automated fashion. In this white paper we presented one possible approach to automation -- artificial neural networks -- that has a unique capability of processing hundreds of thousands of LCs in a matter of minutes. Suitable training data-sets will have to be created that would allow for a wide range of light curve types to be automatically processed. The community will need to invest significant effort to further develop automation methods and update the current tools to cope with this challenge.  In addition, greater attention needs to be paid to intelligent components, such as expert systems, to insure appropriate flow down the data pipeline.

\newpage
\section{Recommendations}

\bf Our recommendations to the Decadal Survey 2010 regarding actions that need to be taken in order to address the challenges pointed out above are:

\begin{itemize}
\item form a dedicated center (such as MAST or HEASARC) that would address the issues of data analysis automation; such a center would employ 3-5 FTE in software engineering and theoretical scientific modeling;
\item form a narrowly focused IAU commission that would steer community efforts -- i.e.~the National Virtual Observatory (NVO) interface, the choice of a programming language (i.e.~python), specifications for application deployment, a well-defined suite of test cases, etc;
\item organize regular workshops and splinter meetings at the AAS and IAU symposia to address the application of Artificial Intelligence and other fully automated methods in astronomy data mining;
\item secure adequate funding through NSF/NASA for technology research and implementation through specialized calls for proposals.
\end{itemize}

\rm

\bigskip

\bibliography{Prsa_Survey_Analysis}

\end{document}